\documentstyle[preprint,version2,aps]{revtex}
\draft
\begin{document}
\begin{title}
\centerline{The Metal-Insulator Transition in the}
\centerline{ Hubbard Model at Zero Temperature II}
\end{title}

\author{Marcelo J. Rozenberg, Goetz Moeller, Gabriel Kotliar}
\begin{instit} Serin Physics Laboratory, Rutgers University,
Piscataway, NJ 08855-0849, USA
\end{instit}
\begin{abstract}
We study the metal-to-insulator transition of the Hubbard model at zero
temperatures in infinite dimensions.
The coexistence of metallic and insulating solutions for a finite
range of the interaction is established.
It is shown that the metallic solution is lower in energy for
any interaction in the coexistence region and that
the transition is of second order.

\end{abstract}
\newpage

{\em Introduction}
The  correlation induced metal-insulator transition (Mott-Hubbard transition)
is one of the prime examples in which
strong correlations dominate the low-energy behavior of a physical
system. A theoretical treatment of the problem requires an
approach which is non-perturbative in the interaction.
Recently, new insights into the problem were gained using
the limit of infinite dimensionality \cite{voll,MeVo}.
It allows for a mapping of a variety of lattice models onto
impurity problems in a self-consistently determined bath
\cite{GKS,GeKo} and is therefore a natural way to formulate a mean-field
theory of itinerant systems.
While being simpler than the original problem,
the resulting mean-field theory remains a formidable many-body problem
which has to be solved using  numerical methods.
Recently the Hubbard model has been investigated by several groups using
Quantum Monte Carlo (QMC) simulations and self-consistent perturbation
theory (PT).\cite{jarrel,RZK,GK,ZRK}
While a  combination of both methods established the existence
of a Mott-Hubbard  transition at
a finite value of the interaction  $U$ in the paramagnetic
phase of the Hubbard model at half-filling,
important questions regarding the nature of the transition remain unsolved.

In a previous work \cite{long}, the
coexistence  of metallic and insulating solutions
over a finite range of values of $U$ has been demonstrated.
While the metallic solution disappears continuously at a value
$U_{c2}$, the insulating solution disappears abruptly at a
value $U_{c1}< U_{c2}$.
At finite temperature, the difference between the free energy  of the
solutions is dominated by the entropy term. The large entropy, which
is a result of the degeneracy of the ground state in the insulating
case, made it possible to unambiguously determine the existence of a
first order transition line close to $U_{c1}(T)$.
As the temperature is reduced, the free energy approaches the energy,
therefore  an accurate evaluation of the energy is necessary.
Depending on which solution is lower in
energy two very different scenarios may take place: If $E_{Ins} <
E_{Met}$, the transition will be close to $U_{c1}$
and the sudden destruction of the metallic state implies a first-order
transition even at $T=0$. On the other hand, in the case $E_{Met} < E_{Ins}$,
the metallic solution continuously merges  with the insulating one at
$U_{c2}$, and the quasiparticles display a diverging renormalized
mass.\cite{long}

While the limit $T = 0$ cannot be attained by QMC simulations, within the
second-order perturbative approach the energies of the two solutions are almost
degenerate, making the consideration of higher-order corrections necessary.
An alternative numerical approach to the problem was introduced recently:
While the large $d$ mean field equations are {\it functional
equations} for the Green function $G(i \omega_n)$, an approximation can be
obtain by modeling $G(i\omega_n)$ using a finite number $N$ of
parameters, which
reduces the functional equations to non-linear algebraic equations in
$N$ unknowns. Following this idea, two different parameterizations
were introduced.\cite{CaKr,QRKR} Both take advantage of
a mapping of the lattice problem onto an Anderson impurity model with a
self-consistently determined bath. The $N$ parameters that model
$G(i\omega_n)$ define the hopping amplitudes and energies of the
effective electron orbitals of the bath, as will be discussed in detail
in next section.
The resulting problem can then be solved at $T=0$ by exact
diagonalization of the effective Hamiltonian. This is followed by the new
determination of the set of parameters, and the procedure is iterated
until convergence is attained. The method is thus non-perturbative in nature
and overcomes the problems of both
QMC and PT, allowing for an accurate evaluation of the energies at $T=0$.

In this paper we apply  this approach to the study of the Hubbard
model. We establish the coexistence of metallic and
insulating solutions over a finite range of the interaction parameter
$U$ and show that at $T=0$  the metallic solution has lower  energy
than the insulating one,
implying  that the metal-insulator transition in the Hubbard
model with semicircular density of states is of second order.
This justifies a posteriori the relevance of the earlier studies
\cite{ZRK}  of this quantum critical point which captures the essence
of the Brinkman-Rice transition.

{\em Methodology}
In the limit of infinite dimensionality the Hubbard model, described
by the Hamiltonian
\begin{eqnarray}
H= -\sum_{<i,j>} (t_{ij}+\mu) c_{i,\sigma}^{\dagger} c_{j,\sigma}
 + U \sum_i (n_{i \uparrow}-\frac{1}{2}) (n_{i\downarrow}-\frac{1}{2}),
\label{HubHam}
\end{eqnarray}
can be reduced to an effective impurity problem, supplemented by a
self-consistency condition.\cite{GeKo}
As in the previous work we focus on a Bethe lattice of infinite
connectivity $m$, which in the non-interacting limit
corresponds to a semicircular density of states of width $4t$,
where the hopping parameter $t$ is rescaled
in the usual way as $t \rightarrow \frac{t}{\sqrt{m}}$.
Integrating out the degrees of freedom other than the origin, one
obtains an effective local action of the form

\begin{eqnarray}
S_{eff}[c, c^{\dagger}] =&& \sum_{\sigma} \int d\tau d\tau'
c^{\dagger}_{\sigma}(\tau )
G_0^{-1} (\tau-\tau') c_{\sigma}(\tau')\nonumber\\
&&+ U \int_0^{\beta}d\tau  (n_{i\uparrow}(\tau)-\frac{1}{2})
(n_{i\downarrow}(\tau)-\frac{1}{2}) .
\label{action}
\end{eqnarray}
In the following we
focus on the paramagnetic solution at half filling $(\mu = 0)$.
The self-consistency condition then reads
$G_0^{-1}(i\omega_n) =  i\omega_n - t^2 G(i\omega_n)$
where
$G(i\omega_n)=-\int_0^{\beta} {\rm e}^{i \omega_n \tau} < T_\tau c(
\tau) c^{\dagger}(0) >_{S_{eff}} $ is  the
${local}$ Green function of the Hubbard model once self-consistency is
attained.
As shown in Ref. \cite{GeKo} an action of the same form can be
obtained from an Anderson impurity
model by integrating out the conduction electrons. Note that
the self-consistency condition implies that
the role of the hybridization function is played by the local Green
function itself.
The iterative solution now proceeds as follows:  $G(i\omega_n)$ is
modeled by a finite set of parameters. In terms of the impurity problem, this
represents an effective bath for the impurity with a finite number of poles.
This effective impurity model is then solved
by exact diagonalization and a new $G(i \omega_n)$ is calculated.
A new set of parameters is then obtained from $G(i\omega_n)$
by approximating it by a function with a number of poles equal to
the number of sites
in the bath (this number is in general smaller than the number of poles of
$G(i\omega_n)$). Note that this represents  a further approximation
of the method (beyond the effective Hamiltonian being finite).
The whole process is iterated until convergence of the parameters is
achieved.

Exploiting these features, two new similar
algorithms were proposed recently \cite{CaKr,QRKR}, differing
basically in the way the new set of parameters is obtained, that is, how the
$G(i\omega_n)$ is parametrized by a smaller number of poles.
We will consider both schemes and comment on their respective advantages
and limitations.

As  mentioned, the number of poles of $G(i \omega_n)$ is in general
larger that the number of sites in the bath, therefore this approximation
is an essential ingredient of the scheme. Caffarel and Krauth \cite{CaKr}
proposed to obtain the new parameters by a
$\chi^{2}$ fit of $G( i\omega_n)$.
Starting with an Anderson Hamiltonian of the form

\begin{eqnarray}
H &=& \sum_{\alpha,\sigma} \epsilon_\alpha a_{\alpha\sigma}^{\dagger}
a_{\alpha\sigma}
  + \sum_{\alpha,\sigma}(       V_\alpha a_{\alpha\sigma}^{\dagger}
c_{ \sigma} + h.c.)\nonumber\\
  &&+ U (n_{c \uparrow}-\frac{1}{2})( n_{c \downarrow}-\frac{1}{2})
\end{eqnarray}
the self-consistency condition becomes
$t^2 G(i\omega) = \sum_{\alpha=1}^{N_s} \frac{V_\alpha^2}
{i \omega_n - \epsilon_\alpha}$.
We thus have to minimize
\begin{equation}
\chi^2 = \sum_{i \omega_n}^{N_{\Omega}} |G(i \omega_n) -
\sum_{\alpha=1}^{N_{site}}
\frac{V_\alpha^2}{i \omega_n - \epsilon_\alpha} |^2
\end{equation}
where we sum over frequencies  $\omega_n= (2n+1)\pi T$ with small fictitious
temperature ( $T=.001$) and large cutoff $N_{\Omega} \Delta \omega
\approx 2 U$, to obtain the new set of parameters $V_{\alpha}$ and
$\epsilon_{\alpha}$. Note that this Hamiltonian effectively
describes an impurity surrounded by a ``star'' of bath electrons.

An alternative route was introduced in the context of an
extended Hubbard model\cite{QRKR}. This procedure takes advantage of
the fact that the Green function $G(z)$ can be decomposed into
``particle'' and  ``hole'' contributions as
$G(z)=G^>(z)+G^<(z)$ with
$G^>(z)=<0|c \frac{1}{z-(H-E_0)} c^{\dagger} |0>$ and
$G^<(z)=<0|c^{\dagger}\frac{1}{z+(H-E_0)} c |0>. $

The respective contributions can be obtained from a continued fraction
expansion as
\begin{equation}
<f_0^{>/<}|\frac{1}{\omega\mp(H-E_0)}|f_0^{>/<} > =
\frac{<f_0^{>/<}|f_0^{>/<}>}{z-a_0^{>/<}
-\frac{b_1^{>/<2}}{z-a_1^{>/<}-\frac{b_2^{>/<2}}{z-a_2^{>/<}-...}}}
\end{equation}
where $|f_0^>> = f^{\dagger} |gs>$, $|f_0^<> = f |gs>$
and $|f_{n+1}> = H |f_n>-a_n |f_n> -b_n^2 |f_{n-1}>$,
$a_n=<f_n|H|f_n>$, $b_n^2 = \frac{<f_n|f_n>}{<f_{n-1}|f_{n-1}>}$, $b_0=0$.
This implies that
$G^>$ and $G^<$ can be regarded as resulting from a Hamiltonian
describing an impurity coupled to  two chains
with site energies  $a_n^{>/<}$ and hopping amplitudes $b_n^{>/<}$.
Again  the number of poles in the Green function  is in general larger
than the number of sites of the Hamiltonian and
in order to close the self-consistency, the continued fraction
expansion has to be truncated.
The approximation in this scheme relies on the fact that the continued
fraction  representation captures exactly the moments of the energy of
the Hamiltonian, up to the order retained in  the continued fraction.
It can thus be thought of as a moment by moment fitting.
This scheme has the numerical advantage that it avoids
the multidimensional fit of the Green function, but the
disadvantage that it can be implemented practically only
in the case of a semi-circular density of states.
In the metallic case an explicit extra site
at the Fermi energy is introduced in order to better represent the low
frequency region  and, more importantly, to allow us to  feed-back a
metallic bath. The hopping parameter to this extra site is calculated by
a single parameter minimization of the expression
\begin{eqnarray}
 \chi^2(\alpha) = \sum_{i \omega_{n L}}^{i\omega_{n H}}
|G_A(i \omega_n,\alpha) - G(i \omega)|^2
\end{eqnarray}
where now
$G_A(i \omega_n,\alpha) = \frac{\alpha}{i\omega_n} + (1-\alpha)
G_{N_C}(i\omega_n)$.
$G_{N_{C}}$ is the truncated Green function to length
$N_{C}=N_{Site}/2$  and
$\omega_L$ and $\omega_H$ are low and high energy cut-offs defined by
the lowest poles of $G$ and $G_{N_{C}}$, respectively.
In this case the moments will be modified by a small factor
$(\alpha)$ which decreases as the system size is increased.

The effective Anderson model therefore reads
\begin{eqnarray}
H &=& \sum_{\sigma}\sum_{\rho=>,<}(\sum_{\alpha=1}^{N_C-1}
 a_{\alpha}^{\rho} c_{\alpha\sigma}^{\rho\dagger} c_{\alpha\sigma}^{\rho}
+b_0^{\rho}( c_{\sigma}^{\rho\dagger} f_{\sigma} + h.c.)\nonumber\\
&&+\sum_{\alpha=1}^{N_{C}-2}
( b_\alpha^{\rho} c_{\alpha\sigma}^{{\rho}\dagger} c_{ \alpha+1\sigma}^{\rho}
  +h.c. ))
  + U (n_{f \uparrow}-\frac{1}{2}) (n_{f \downarrow}-\frac{1}{2}) \nonumber\\
&&  +\sum_{\sigma} b_0 (f_{\sigma}^{\dagger} c_{0\sigma} + h.c.).
\end{eqnarray}

In both schemes, groundstate wavefunction and groundstate energy
of the Anderson Hamiltonian
are determined by exact diagonalization (up to six sites) and
the modified Lanczos technique\cite{elbio}. Systems of
up to ten sites can be handled on a workstation.
The zero temperature Green function of the local site is finally obtained
from a continued fraction expansion using the recursion method discussed
above.

As mentioned in the introduction, a further
advantage of  the  formulation of the problem in terms of
an Anderson impurity model is the fact that
the energy of the Hubbard model can be obtained directly without
frequency summations using  Anderson model relations.
The kinetic energy per site of the Hubbard model  is given
as $T=\frac{2}{\beta N}
\sum_{<j,k>}\sum_{i \omega_n} t G_{jk}(i \omega_n) e^{i \omega_n
0^+}$. Taking the limit of infinite coordination number this reduces
to $T=\frac{2 t^2}{\beta} \sum_{i \omega_n}
G(i\omega_n)^2 e^{i \omega_n 0^+}$. Using  the self-consistency
condition as well as the  the fact that in the Anderson model
$\frac{2}{\beta}\sum_{i\omega_n}\sum_{\alpha \sigma} \frac{V_{\alpha}^2}{i
\omega_n-\epsilon_{\alpha}}  <f_{\sigma} (i \omega_n)
f_{\sigma}^{\dagger} (i \omega_n) > \nonumber\\
=\sum_{\alpha
\sigma} V_{\alpha} <f_{\sigma}^{\dagger} c_{\alpha \sigma} + h.c. >$
we  obtain
\begin{equation}
T  =  \sum_{\alpha \sigma} V_{\alpha} Re <0|
f_{\sigma}^{\dagger} c_{\alpha \sigma} |0>,
\end{equation}
where $\alpha$ labels the sites neighboring the impurity.
The potential energy of the Hubbard model is obtained as
\begin{equation}
V = U <0| n_{f \uparrow}  n_{f \downarrow} |0>.
\end{equation}

{\em Results}
In our analysis we have focused on two major aspects:  the determination
of a region where two solutions are allowed, and the
resolution of controversy  regarding the lowest energy solution.
The two approaches considered yield a consistent picture of
the transition.
We are able to obtain converged metallic and insulating solutions
for a finite range of the interaction $U$ within both schemes.
We further demonstrate that the metallic solution is lower in energy
in the whole coexistence region.
The energy
difference between the solutions goes to zero as $U_{c2}$ is
approached, implying that the transition can be classified as
second order. This should be contrasted with the results
from second-order perturbation theory, where the two solutions were
found to cross in energy at an intermediate value of the interaction $U$.
A point worth noticing (as was already observed within
the perturbative approach) is that the
energy difference between the solutions is much smaller than any
energy  scale of the problem. This is due to an almost perfect
compensation of the gain in delocalization (kinetic) energy, by the
loss of energy through double occupancy (potential energy), in the
metallic state compared to the insulator.

Metallic and insulating
solutions for  $U=2.7$ inside the coexistence
region are respectively shown in Fig.\ref{Fig1} and Fig.\ref{Fig2}
(the half-bandwidth $2t$ is
set equal to unity). In the first figure the Green function displays a narrow
resonance at low frequency (note that the pinning condition at $\omega=0$ is
fulfilled \cite{MH}), while the insulator in the second case merely
consists of high energy features (upper and lower Hubbard bands).
The figures also illustrate the consistency of the two schemes
considered here. In both, the metallic and insulating,  cases the agreement is
very good. We also find that the results
of both methods for the single particle Green
function on the imaginary axis compare very well with
the second-order perturbative calculation\cite{ZRK} and QMC\cite{RZK,GK}
(the latter is discussed in Ref.\cite{CaKr}).

The kinetic, potential and total energies for the
two solutions in the coexistence region are displayed in Fig.\ref{Fig2}.
An interesting feature is the already mentioned almost perfect
cancellation of delocalization and double occupancy energy.
Another important observation is
that while a finite size effect is apparent in the results for the
kinetic and   potential energy, the convergence of the
total energy is much faster\cite{book}.
A few runs for a ten site system show almost no difference to
the results for eight sites.

The energy difference of the two solutions
is shown in the inset of Fig.\ref{Fig2}. As the critical
point $U_{c2}$ is approached from below, the finite size effects
become relevant for $U \approx 2.8$. This limitation of the scheme is
due to the fact that as the low energy scale associated with
the quasiparticle peak goes to zero close to the transition,
the discrete nature of the approximation
starts playing an important role and the Kondo resonance  is
represented by  only a single pole.

The smallness  of the difference in energy between the metal and the
insulator can be understood from
the picture of a second-order critical point where the
metallic and insulating solutions merge with a vanishing scale
$\Delta \sim U_{c2}-U$.   The problem can be formulated from a variational
point of view, with the free energy $F$ becoming an extremum at the metallic
and insulating solutions, i.e.,
$\frac{\delta F}{\delta G_M}= \frac{\delta F}{\delta G_I}=0$.
Since the two solutions merge at the point
$U_{C2}$, $F$ can be expanded in power series of $G_M-G_I$ as
\begin{equation}
F_M-F_I=\frac{\delta^2 F}{\delta G^2}(G_M-G_I)^2.
\end{equation}
As the difference between the metallic and the insulating solution
is parameterized by $\Delta$, and the second derivative vanishes
at the critical point as $\Delta$, it follows that the energy
difference goes to zero as $\Delta^3$.
The critical region cannot be accessed by the present method.
In order to capture the vanishing energy scale, a higher resolution (i.e.
an effective bath with more sites) is needed.

Finally, we would like to comment on the disappearance of the
insulating solution  at $U_{c1}$. From the
"two chain" scheme, the insulating solution is found to persist all the way
down until the gap closes. This  differs from the results of perturbation
theory and  resembles the Hubbard III scenario
for the destruction of the insulating state.\cite{HIII,gros}
In the case of the "star
configuration", while a converged insulating solution can be obtained at
values of the interaction $U$ well below $U_{c2}$, the question of the closing
of the gap cannot be answered conclusively.

{\em Conclusion}
We have resolved the standing questions regarding the metal-to-insulator
transition in the Hubbard model in infinite dimensions, using a
powerful algorithm to obtain Green functions at zero temperature\cite{dyncor}.
We were able to demonstrate the existence of a region in which
metallic and insulating solutions coexist, which is in agreement with
previous results, and showed that the metallic solution is always lower
in energy. This implies that while at finite temperature the transition
is first order, it becomes second-order at $T=0$, similar to
the work of Brinkman and Rice in the context of the Gutzwiller
approximation  \cite{BR} \cite{ZRK}.
Since the method presented is  very general as well as simple,
especially when compared to Monte Carlo simulations,
it is an appealing approach to the study of strongly correlated
electron systems.

\acknowledgements{We would like to thank E. Dagotto, V.
Dobrosavljevi\'{c}, A. Moreno, A. Ruckenstein and Q. Si
for stimulating discussions.
This work was supported by the NSF under grant DMR-92-24000.}

\figure
{Comparison of the metallic and insulating Matsubara Green functions for
$U=2.7$, as obtained from the two variations of the algorithm.
Full line "star geometry" and dotted line "two chain geometry" (10
sites for the metallic case and 8 sites for the insulating)
\label{Fig1}}

\figure
{Kinetic, potential and total energy for the
metallic and insulating solutions in the
coexistence region. Difference between the metallic and insulating
solution (inset). From the "two chain geometry".
\label{Fig2}}


\begin{references}
\bibitem{dyncor} Note that this can also be used to obtain dynamical
correlation functions like the spin-spin correlation function.
\end{references}
\end{document}